\documentclass[10pt,journal]{IEEEtran}
\IEEEoverridecommandlockouts
\usepackage{fancyhdr}
\usepackage{multirow}
\usepackage{mdwlist}
\usepackage{amsmath}
\usepackage{amssymb}
\usepackage{latexsym}
\usepackage{CJK}
\usepackage{subfigure}
\usepackage{indentfirst}
\usepackage{ url}
\usepackage{graphicx}
\usepackage{subfigure}
\usepackage{epstopdf}
\usepackage{longtable}
\usepackage{array}
\usepackage[thmmarks,amsmath]{ntheorem}
\usepackage{diagbox}[2011/11/22]
\usepackage{textcomp,booktabs}
\usepackage[usenames,dvipsnames]{color}
\usepackage{colortbl}
\usepackage{breqn}
\usepackage{flushend}
\usepackage{mathrsfs}
\usepackage{stfloats}
\usepackage{algorithmic}
\usepackage{algorithm}
\usepackage{amsmath}
\usepackage{color}
\usepackage{engord}
\usepackage{makecell}
\usepackage{tcolorbox}

\begin{document}

\title{Accuracy and Security-Guaranteed Participant Selection and Beamforming Design for RIS-Assisted Federated Learning}

\author{
Mengru Wu, Yu Gao, Weidang Lu, Huimei Han, Lei Sun, and Wanli Ni

\thanks{\vspace{0.5em}}
\thanks{Mengru Wu, Yu Gao, Weidang Lu and Huimei Han are with the College of Information Engineering, Zhejiang University of Technology, Hangzhou 310023, China (e-mail: wumengru@zjut.edu.cn; gaoyu0806@foxmail.com; luweid@zjut.edu.cn; hmhan1215@zjut.edu.cn).}
\thanks{Lei Sun is with the School of Automation and Electrical Engineering, University of Science and Technology Beijing, Beijing 100083, China
 (e-mail: sun_lei@ustb.edu.cn).}
\thanks{Wanli Ni is with the Department of Electronic Engineering, Tsinghua University, Beijing 100084, China
	(e-mail: niwanli@tsinghua.edu.cn).}}

\maketitle

\begin{abstract}
Federated learning (FL) has emerged as an effective approach for training neural network models without requiring the sharing of participants' raw data, thereby addressing data privacy concerns. In this paper, we propose a reconfigurable intelligent surface (RIS)-assisted FL framework in the presence of eavesdropping, where partial edge devices are selected to participate in the FL training process. In contrast, the remaining devices serve as cooperative jammers by transmitting jamming signals to disrupt eavesdropping. We aim to minimize the training latency in each FL round by jointly optimizing participant selection, bandwidth allocation, and RIS beamforming design, subject to the convergence accuracy of FL and the secure uploading requirements. To solve the resulting mixed-integer nonlinear programming problem, we propose a twin delayed deep deterministic policy gradient (TD3) algorithm. Simulation results demonstrate that the proposed scheme reduces the FL training latency by approximately 27$\%$ compared to baselines.
\end{abstract}

\IEEEpeerreviewmaketitle

\section{Introduction}
With the rapid development of the Internet of Things, vast amounts of data are generated by edge devices. By leveraging the massive data, neural network models can be trained to support intelligent services [1]. However, traditional machine learning (ML) approaches involve significant communication overheads and raise data privacy concerns due to the transmission of raw data to an edge server. Federated Learning (FL) has been proposed as an effective solution to address these issues. Specifically, FL enables participants to train neural network models locally and upload only model parameters to the edge server, thereby avoiding sharing raw data [2].

Given the challenges posed by undesirable wireless fading, reconfigurable intelligent surfaces (RISs) have emerged as a viable solution to assist the interactions between an FL server and edge devices [3]. By modifying the reflecting elements of a RIS, the received signal power at the FL server can be enhanced [4]. In [5], the authors developed a gradient quantization system, where a unified optimization problem was examined by simultaneously optimizing power allocation, sub-band assignment, and RIS beamforming. Driven by the benefits of the RIS technology, several studies have employed the RIS to assist FL. Specifically, the authors in [6] employed a RIS to enhance transmission efficiency. They also jointly optimized RIS phase shifts, communication resource, transmit power, and local computing frequencies to minimize the total training latency of FL. In [7], the authors studied a RIS-aided FL system with statistical channel state information, where RIS beamforming and bandwidth allocation were jointly designed to minimize outage probability.

In light of the inherent security vulnerabilities of wireless channels, uploaded model parameters or gradients are susceptible to eavesdropping. Physical layer security (PLS) has attracted extensive attention as a key technique for securing communications. PLS utilizes wireless channel characteristics to enlarge the disparity between legitimate and eavesdropping channels, thereby preventing uploaded model parameters from eavesdropping [8]. Therefore, it is worthwhile to investigate PLS to ensure the security in FL systems. The authors in [9] developed an unmanned aerial vehicle (UAV)-enabled FL architecture, in which a UAV not only acted as a server for FL but also transmitted artificial noise (AN) to disturb eavesdropping. In [10], participants for training FL were randomly selected, while other devices sent jamming signals to impede eavesdropping. Also, the authors in [10] jointly optimized time allocation, jamming power, and FL local iterations to minimize devices' energy consumption.

While prior works [3]-[7] have utilized the RIS technology to enhance model uploading efficiency in FL systems, they have overlooked the secure transmission of model parameters. Furthermore, [8]-[10] have demonstrated the advantages of PLS for safeguarding uploaded model parameters in the presence of eavesdropping. Nevertheless, these works have not considered deploying a RIS to improve the quality of wireless links in FL networks. Besides, the previous works in [3]-[10] have not explored the convergence accuracy of FL training, which is vital for ensuring the stability and interpretability of training a global model in distributed wireless environments.

Motivated by the aforementioned discussions, this paper presents a RIS-assisted FL framework by guaranteeing the convergence accuracy and the transmission security of model parameters in FL systems. Specifically, a RIS is used to assist the uploading of model parameters from edge devices to an edge server in the presence of an eavesdropper. A subset of edge devices with powerful computational capabilities or good communication conditions are selected as participants in the training process, while the remaining devices act as cooperative jammers to interfere with eavesdropping. To minimize the training latency of each round in FL, we jointly optimize participant selection, bandwidth allocation, and RIS beamforming design while ensuring the convergence accuracy of FL and secrecy rate requirements.

To summarize, the main contributions of this paper are listed as follows:
\begin{itemize}
\item\textit{Accuracy and Security-Guaranteed FL:} We propose a novel scheme that ensures the convergence accuracy and the secure transmission of model parameters in FL systems in the presence of an eavesdropper. We first analyze the impact of participant selection on the convergence accuracy of FL by deriving a convergence upper bound of the average global gradient norm. Then, we formulate a training latency minimization problem by jointly optimizing participant selection, bandwidth allocation, and RIS beamforming subject to the convergence accuracy and transmission security constraints.
\item\textit{TD3 Algorithm Design:} To address the aforementioned training latency minimization problem, the formulated optimization problem is modeled as a Markov Decision Process (MDP) to describe the interactions between the BS and the RIS-assisted FL environment. Then, a twin delayed deep deterministic policy gradient (TD3)- based participant selection and beamforming design (PSBD) algorithm is developed for participant selection, bandwidth allocation, and beamforming design, thereby efficiently achieving joint decision-making for training latency minimization in the dynamic environment.
\item\textit{Performance Evaluation:} We perform simulations to evaluate the performance of the proposed accuracy and security-guaranteed FL scheme. The simulation results show the superiority of the proposed TD3 algorithm in convergence compared to the deep deterministic policy gradient (DDPG) algorithm. Additionally, simulation results illustrate that the proposed scheme reduces FL training latency by approximately 27$\%$ compared to baselines.
\end{itemize}

\section{System Model}
As depicted in Figure 1,  we consider a RIS-assisted FL system consisting of a BS equipped with an edge server, $K$ single-antenna edge devices denoted by ${\cal K} = \!\{1,...,K\}$ , a RIS equipped with $M$ passive reflecting elements indexed by ${\cal M} = \!\{1,...,M\}$, and an eavesdropper (Eve). In this system, edge devices and the BS collaboratively train a neural network model with the assistance of the RIS. The RIS can secure local model uploading in the presence of the Eve by flexibly adjusting its phase shifts of reflecting elements. Each edge device $k$ possesses its local dataset
	${{\cal D}_k} = \{ ({\textbf{\textit{u}}_{ki}},{\textit{v}_{ki}})\} _{i = 1}^{{D_k}}$, where $\textbf{\textit{u}}_{ki}$  is the $i$-th sample vector containing features, $\textit{v}_{ki}$ is the corresponding ground-truth label for the sample ${\textbf{\textit{u}}}_{ki}$, and ${D_k}$ is the number of training samples of edge device $k$. Thus, we obtain a global dataset
	$\{ {{\cal D}_k}\} _{k = 1}^K$ that encompasses
	$\left| {\cal D} \right| = \sum\nolimits_{k = 1}^K {\left| {{{\cal D}_k}} \right|} $ training samples. To alleviate communication burdens, a subset of edge devices with powerful computational capabilities or good communication conditions are selected as participants in the training process, while the remaining devices act as cooperative jammers to interfere with eavesdropping. Let
	${x_k}\in \{0,1\}$ denote the participant selection indicator of device $k$. Specifically,
	${x_k} = 1$ means that device $k$ is selected to upload its local model to the BS, while ${x_k} = 0$ indicates that device $k$ actively sends jamming signals as a jammer. Define
	${{\cal A}} = \{  k \in {\cal K} \mid {{x_k}} = 1 \}$ as the set of participants and
	${{\cal A}'} = \{  k \in {\cal K} \mid {{x_k}} = 0 \}$ as the set of jammers, and ${{\cal A}} \cup {{\cal A}'} = {\cal K}$. For the sake of distinction, we employ $\hat k$ and $\bar k$ as the indicators for a participant and a jammer, respectively.


\begin{figure}[htbp]
\centering
\includegraphics[width=0.48\textwidth]{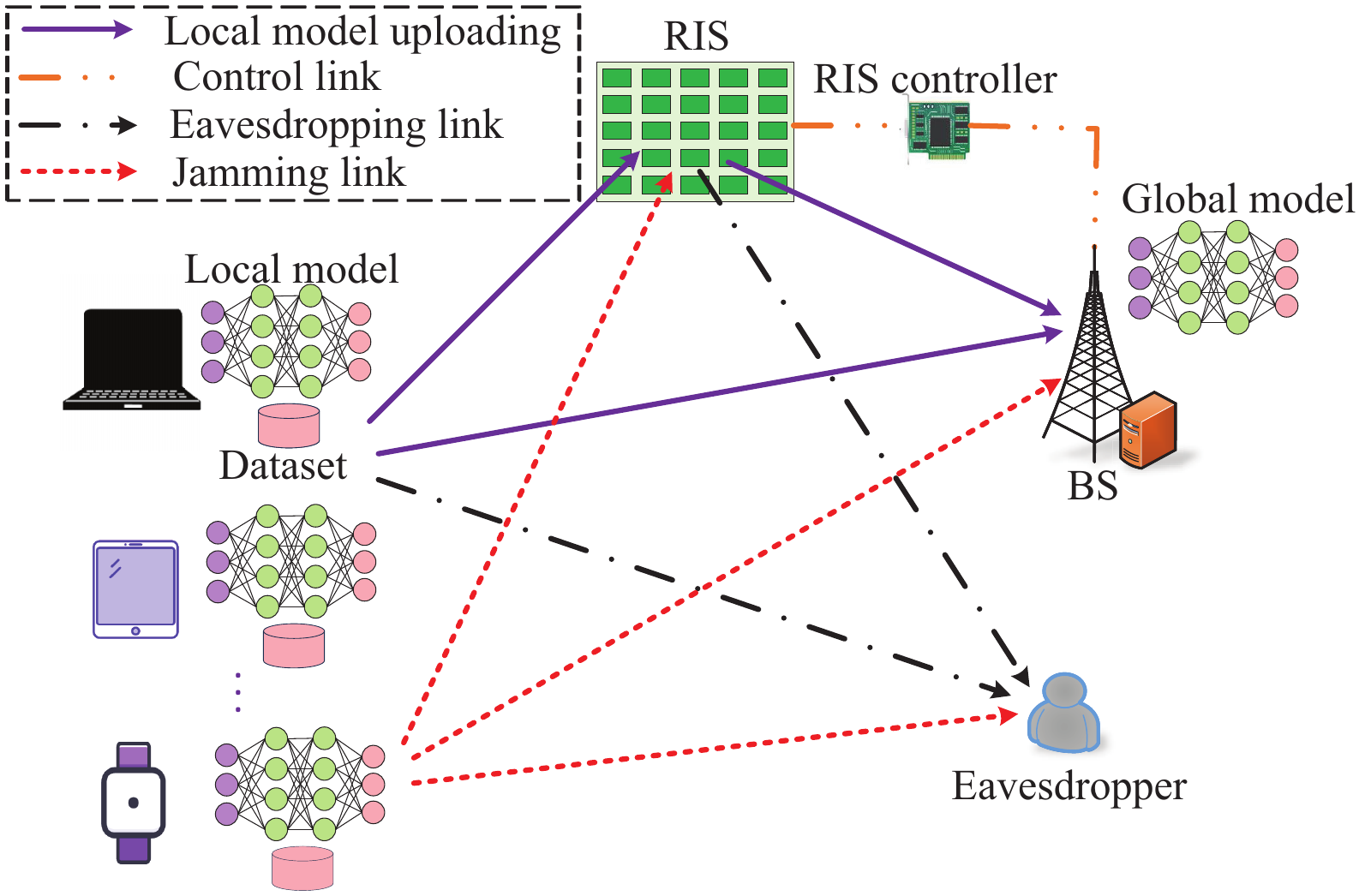}
\caption{An illustration of a RIS-assisted FL system.}
\end{figure}

\subsection{FL Training Process}
Let $\textbf{\textit{w}}\in \mathbb{R} ^{d} $ and $\textbf{\textit{w}}_{k} \in \mathbb{R} ^{d} $ represent $d$-dimensional global model parameters and local model parameters of device $k$, respectively. The loss function of device $k$ that evaluates the training error on the dataset ${{\cal D}_k}$ is expressed as [6]

	\begin{equation}
		{F_k}(\textbf{\textit{w}}) = \frac{1}{{{D_k}}}\sum\limits_{i = 1}^{{D_k}} {f(\textbf{\textit{w}};{\textbf{\textit{u}}_{ki}},{\textit{v}_{ki}})}, \quad \forall k \in {\cal K},
    \label{eq:1}
	\end{equation}
where ${f(\textbf{\textit{w}};{\textbf{\textit{u}}_{ki}},{\textit{v}_{ki}})}$ represents the sample-wise loss function. Therefore, the global loss function at the BS is defined as

	\begin{equation}
		F(\textbf{\textit{w}}) = \frac{1}{D}\sum\limits_{k \in {\cal K}} {{D_k}{F_k}(\textbf{\textit{w}})},
    \label{eq:2}
	\end{equation}
	where
	$D = \sum\nolimits_{k = 1}^K {{D_k}} $. The learning process can be viewed as solving the following optimization problem
	\begin{equation}
\vspace{-0.3em}
		{\textbf{\textit{w}}^*} = \mathop {\arg \min }\limits_{\textbf{\textit{w}} \in {\mathbb{R}^d}} F(\textbf{\textit{w}}),
    \label{eq:3}
	\end{equation}
	where ${\textbf{\textit{w}}^*} = \textbf{\textit{w}} = {\textbf{\textit{w}}_1} = ... = {\textbf{\textit{w}}_K}$. To address the above problem, the BS determines the set of edge devices taking part in the current round and broadcasts the global model ${\textbf{\textit{w}}^{t-1}}$  at round $t-1$ to all the participants. Upon receiving the global model ${\textbf{\textit{w}}^{t-1}}$, each participant $\hat k \in{{\cal A}}$ updates its local model by employing the gradient descent algorithm on its local dataset as

	\begin{equation}
		\textbf{\textit{g}}_k^t = \frac{1}{{{D_k}}}\sum\limits_{i = 1}^{{D_k}} {\nabla f({\textbf{\textit{w}}_k^t};{\textbf{\textit{u}}_{ki}},{\textit{v}_{ki}})}, \quad \forall k \in {\cal K},
    \label{eq:4}
	\end{equation}
    \begin{equation}
	\textbf{\textit{w}}_k^t = {\textbf{\textit{w}}^{t - 1}} - \alpha \textbf{\textit{g}}_k^t, \quad \forall k \in {\cal K},
    \label{eq5}
	\end{equation}
	where $\nabla$ indicates the gradient operator and $\alpha$ is the learning rate. Subsequently, participant $\hat k$ transmits the latest model to the BS. The BS carries out global model aggregation as

	\begin{equation}
     {\textbf{\textit{w}}^t} = \frac{{\sum\nolimits_{k = 1}^K {{x_k}\textbf{\textit{w}}_k^t} }}{{\sum\nolimits_{k = 1}^K {{x_k}} }},
     \label{eq6}
	\end{equation}
where ${\textbf{\textit{w}}^t}$ is the updated global model at round $t$. FL attains convergence by iterating the aforementioned steps until the accuracy of the global model reaches the desired threshold.

\subsection{Computation and Communication Models}
\subsubsection{Local Training}
We denote ${c_{\hat k}}$ as the required number of central processing unit (CPU) cycles for executing one data sample on edge device $\hat k$. Accordingly, the total CPU cycles required for executing all samples on device $\hat k$ is ${c_{\hat k}}{D_{\hat k}}$. By defining ${{f_{\hat k}}}$ as the CPU frequency of edge device $\hat k$, the local training latency of edge device $\hat k$ can be calculated by

	\begin{equation}
		T_{\hat k}^{loc} = \frac{{{c_{\hat k}}{D_{\hat k}}}}{{{f_{\hat k}}}}, \quad \forall  \hat k \in {{\cal A}}.
    \label{eq7}
	\end{equation}

\subsubsection{Local Model Uploading}
We assume that each round of global training can be accomplished within a time slot, and channel conditions remain constant during this time slot. With the aid of the RIS, the received signal power through legitimate links can be substantially increased in comparison to eavesdropping links.
 The achievable rate from participant $ {\hat k} \in {{\cal A}}$ to the BS and the Eve can be respectively expressed as

	\begin{equation}
\vspace{-1em}
R_{\hat{k},b}\!=\! b_{\hat{k}} B \log_2(1\!+\!\frac{P_{\hat{k}}\!\left| {\bf{h}}_{r,b}^H{\bf{\Theta}} {\bf{h}}_{\hat{k},r} \!+\! h_{\hat{k},b} \right|^2}{\sum\limits_{\bar{k} \in {{\cal A}'}}\!\!\! P_{\bar{k}}\!\left| {\bf{h}}_{r,b}^H{\bf{\Theta}} {\bf{h}}_{\bar{k},r} \!+\! h_{\bar{k},b} \right|^2 \!+\! \sigma_b^2}\!),
    \label{eq8}
	\end{equation}

	\begin{equation}
R_{\hat{k},e} \! = \! b_{\hat{k}} B \log_2(1 \! + \! \frac{P_{\hat{k}} \left| {\bf{h}}_{r,e}^H {\bf{\Theta}} {\bf{h}}_{\hat{k},r} \! + \! h_{\hat{k},e} \right|^2}{\sum\limits_{\bar{k} \in {{\cal A}'}}\!\!
\!P_{\bar{k}} \left| {\bf{h}}_{r,e}^H {\bf{\Theta}} {\bf{h}}_{\bar{k},r} \! + \! h_{\bar{k},e} \right|^2 \! + \! \sigma_e^2}\!),
    \label{eq9}
	\end{equation}
	where $b_{\hat k} \in [0,1]$ denotes the bandwidth coefficient assigned to participant $\hat k$, $B$, $P_{\hat k}$,  and $P_{\bar k}$ are the system bandwidth, the transmit power of participant $\hat k$, and the transmit power of jammer $\bar k$. Also, ${{\bf h}_{r,b}}\in \mathbb{C}^{M\times 1}$, ${{\bf{h}}_{\hat k,r}} \in \mathbb{C}^{M\times 1}$, ${{\bf{h}}_{\bar k,r}} \in \mathbb{C}^{M\times 1}$, ${{{h}}_{\hat k,b}}$, ${{{h}}_{\bar k,b}}$, ${{\bf h}_{r,e}}\in \mathbb{C}^{M\times 1}$, ${{{h}}_{\hat k,e}}$, and ${{{h}}_{\bar k,e}}$ respectively indicate the channel coefficients from the RIS to the BS, from participant $\hat k$ to the RIS, from jammer $\bar k$ to the RIS, from participant $\hat k$ to the BS, from jammer $\bar k$ to the BS, from the RIS to the Eve, from participant $\hat k$ to the Eve and from jammer $\bar k$ to the Eve. Besides, $ {\theta_m}$ is the phase shift of the $m$-th reflecting element and $\mathbf{\Theta}  = \mathrm{diag}\{ {e^{j{\theta _1}}},...,{e^{j{\theta _M}}}\} \in \mathbb{C}^{M\times M}$ indicates the passive beamforming matrix, $\sigma _b^2$ and $\sigma _e^2$ stand for the noise power at the BS and the Eve, respectively. On this basis, the secrecy uploading rate from participant $\hat k$ to the BS is written as

	\begin{equation}
		R_{\hat k}^s = \max \{ 0,{R_{\hat k,b}} - {R_{\hat k,e}}\} , \quad \forall \hat k \in {{\cal A}}.
    \label{eq10}
	\end{equation}
	Then, the local model uploading latency of participant $\hat k$ is obtained as

	\begin{equation}
		T_{\hat k}^{tr} = \frac{{S_{\hat k}}}{R_{\hat k,b}},\quad\forall \hat k \in {{\cal A}},
    \label{eq11}
	\end{equation}
where $S_{\hat k}$ represents the size of model parameters.

	It is worth noting that the latency for performing global model aggregation and broadcasting the global model to all participants can be neglected. This is because the BS has powerful computing capability and utilizes the whole downlink bandwidth when broadcasting the global model [11].
Thus, the latency of participant $\hat k$ for one global training iteration is deemed as the sum of the local training latency and the local model uploading latency as

	\begin{equation}
T_{\hat k}^{tot} = T_{\hat k}^{loc} + T_{\hat k}^{tr}, \quad\forall \hat k \in {{\cal A}}.
	\end{equation}

\section{Convergence Analysis of FL and Problem Formulation}
In this section, we evaluate the convergence of the RIS-assisted FL system and examine the training latency minimization for FL.
\subsection{Convergence Analysis of FL}
For the sake of analysis, we adhere to [12] and make the following assumptions for the loss function and  gradients.

\noindent\textbf{Assumption 1.} \textit{(Smoothness). The loss function $F: {\mathbb{R}^d}\to\mathbb{R}$ is $\mu $\textit{-smooth}. Namely, $\forall \textbf{\textit{w}}, \textbf{\textit{w}}' \in \mathbb{R}^d$, we have}

	\begin{equation}
		F({\textbf{\textit{w}}})\le F({\textbf{\textit{w}}}^{'} )+\left \langle \nabla F({\textbf{\textit{w}}}^{'} ) ,{\textbf{\textit{w}}}-{\textbf{\textit{w}}}^{'} \right \rangle +\frac{\mu }{2} \left \| {\textbf{\textit{w}}}-{\textbf{\textit{w}}}^{'}  \right \| _{2}^{2},
	\end{equation}
	\textit{where
	$\left \langle \cdot ,\cdot  \right \rangle $ is the inner product and $\mu$ is the Lipschitz constant.}

\noindent\textbf{Assumption 2.}
\textit{(Unbiased Gradient). Stochastic gradients are unbiased estimates of the local gradients, which can be mathematically represented as}

	\begin{equation}
		\mathbb{E} \left \{ \textbf{\textit{g}}_{k}  \right \} = \nabla F_{k} \left ( {\textbf{\textit{w}}} \right ), \quad \forall k\in \mathcal{K},
	\end{equation}
	\textit{where $\mathbb{E}$ indicates the expectation operator.}

\noindent\textbf{Assumption 3.}
\textit{(Bounded Loss Function). The loss function $F(\textbf{\textit{w}})$ is bounded by ${F^*}$ for any parameter vector $\textbf{\textit{w}}$.}

\noindent\textbf{Assumption 4.}
\textit{(Bounded Gradient Norm). The second moments of stochastic gradients are bounded by a constant $\delta $, i.e.,}

	\begin{equation}
		\mathbb{E} \{{\left\| {{\textbf{\textit{g}}_k}} \right\|^2}\}  \le \delta ,\quad\forall k \in {\cal K},
	\end{equation}
	\textit{where ${\left\| {\bf{a}} \right\|}$ indicates the Euclidean norm of vector $\bf{a}$.}

We utilize the expected average gradient norm to measure the convergence performance of FL, which is commonly employed in analyzing the convergence of non-convex loss function [13]. Given the assumptions stated above, the average gradient norm can be bounded by the average global gradient deviation as follows.

\noindent\textbf{Theorem 1.}
\textit{Given the learning rate satisfying
	$0 < \alpha  \le \frac{1}{\mu }$, the  average norm of the global gradients after
	$\Omega $ rounds can be bounded by}
	\[\frac{1}{{\Omega  + 1}}\sum\limits_{t = 0}^\Omega  {[{{\left\| {\nabla F({\textbf{\textit{w}}^t})} \right\|}^2}]}  \le \frac{{2\mu (F({\textbf{\textit{w}}^0}) - {F^*})}}{{\Omega  + 1}}\]

	\begin{equation}
		+ 2\delta  + \frac{{2\delta ({K} - 2\left| {{{\cal A}}} \right|)}}{{{{\left| {{{\cal A}}} \right|}^2}}}{\sum\limits_{k = 1}^K {x_k} }.
 \label{eq:16}
	\end{equation}

\begin{IEEEproof}
\label{proof1}
Please see Appendix A.
\end{IEEEproof}

\noindent\textbf{Remark 1.}
\textit{Theorem 1 shows a theoretical upper bound for the average norm of the global gradients in FL. It is clear that the right-hand side of (16) comprises three terms, including an initial optimality gap, a bound of the second moments of stochastic gradients, and a term related to participant selection. When the number of communication rounds, the learning rate, the initial model parameter, and $\delta$ are given, the first two terms are constants. The upper bound for the average norm of the global gradients is mainly determined by the third term. Thus, it is important to select a proper subset of edge devices for participating in FL to achieve better training performance [14].}

\subsection{Problem Formulation}

In the RIS-assisted FL system, we aim to minimize the training latency at each FL round by jointly optimizing participant selection, bandwidth allocation, and RIS beamforming design while meeting the convergence and security requirements. Correspondingly, the latency minimization problem can be formulated as

\begin{subequations}\label{eq:main}
\begin{align}
&\mathop {\min }\limits_{\{ {x_{k}},{b_{\hat k}},\mathbf{\Theta}\} } {\mathop {\max }\limits_{\hat k \in \mathcal{A}} T_{\hat k}^{tot}}, \label{eq:18a}\\
\text{s.t.} \quad & \sum\limits_{\hat k \in \mathcal{A}}{b_{\hat k}} \le 1, \label{eq:18b}\\
& \frac{{2\mu (F(\mathbf{w}^0)\!-\!\!{F^*})}}{{\Omega  + 1}}\!+\!2\delta\!+\!\frac{{2\delta ({K}\!-\!\!2\left| \mathcal{A} \right|)}}{{{\left| \mathcal{A} \right|}^2}}\!\!\sum\limits_{k = 1}^K \!{x_k}\!\le\!\epsilon, \label{eq:18c}\\
& R_{\hat k}^s \ge R_{\hat k}^{\min }, \quad \forall \hat k \in \mathcal{A}, \label{eq:18d}\\
& {b_{\hat k}} \ge 0, \quad \forall \hat k \in \mathcal{A}, \label{eq:18e}\\
&{\theta _m}\in[0,2\pi], \label{eq:18f}\\
& x_k \in \{ 0,1\} , \quad \forall k \in \mathcal{K}, \label{eq:18g}
\end{align}
\end{subequations}

	where $R_{\hat k}^{\min }$ is the secrecy uploading rate of participant $\hat k$ and $\epsilon$ means a convergence accuracy threshold. Furthermore, (17b) denotes the bandwidth resource constraint. (17c) ensures that the convergence accuracy of FL exceeds a certain threshold. (17d) indicates that the secrecy uploading rate must surpass a secrecy requirement. (17e)-(17g) signifies that feasible regions of the optimization variables.

\section{Proposed Algorithm}

Due to the non-convex objective function and constraints, the problem in (17) is challenging to solve. Moreover, the discrete variable of participant selection  and the continuous variables regarding bandwidth allocation and RIS beamforming design  are coupled with each other, which further complicates the joint decision-making process. Thus, it is complex to seek real-time optimal solutions for the problem using conventional optimization methods in a dynamic environment [15]. Given the aforementioned considerations, we propose a TD3-based PSBD algorithm to solve the formulated problem efficiently.

\subsection{MDP formulation}

In deep reinforcement learning (DRL), an agent selects an action based on the current state of a complex environment to obtain the optimal reward. As a primary framework for training agents to make sequential decisions, an MDP provides a structured method to model and address decision-making problems encountered in reinforcement learning [16]. Accordingly, the optimization problem outlined in (17)  is modeled as an MDP to describe the interactions between the agent and the environment. The key elements of the MDP model including state space, action space, and reward function, which are detailed as follows.

\noindent\textbf{State space $\mathcal{S}$:} In the RIS-assisted FL environment, the state at each time step ${t}$ contains the total bandwidth allocated to participants and channel state information, which can be expressed as

    \begin{equation}
    s_{t}=\{\sum\limits_{\hat k \in {\cal A}}{b_{\hat k,t}},\{\mathbf{h}_{r,b}^{H}\mathbf{\Theta}_t\mathbf{h}_{\hat k,r}{+}h_{\hat k,b},\,\,\forall \hat k\in\mathcal{A}\}\}.
    \end{equation}
    \noindent\textbf{Action space $\mathcal{A}$:} The decision action of the agent is composed of  participant selection, bandwidth allocation, and beamforming design at the RIS. Thus, the action ${a_{t}\in\mathcal{A}}$ is depicted as

    \begin{equation}
    a_{t}=\{{x_{k,t}},{b_{\hat k,t}},\mathbf{\Theta}_t\}.
    \end{equation}

    \noindent\textbf{Reward function $\mathcal{R}$:} In DRL, a well-designed reward function is essential for guiding the agent toward quickly identifying the optimal policy. In this paper, the reward function consists of the negative value of the optimization objective along with penalties for any constraint violations. Consequently, the reward function is formulated as

\begin{equation}
    r_{t}=-\max T_{\hat{k}\in\mathcal{A}}^{tot}-p_{\text{penalty}},
\end{equation}
where

 \begin{equation}
    p_{\text{penalty}}
	\begin{cases}
0 & \!\!\text{if contraints are satisfied,} \\
   p_1{+}p_2{+}p_3 &\!\! \text{otherwise}, \label{eq:22}
	\end{cases}
    \end{equation}
where  $p_1$, $p_2$, and $p_3$ represent the penalties associated with violating the bandwidth allocation constraint $\eqref{eq:18b}$, the convergence accuracy constraint $\eqref{eq:18c}$, and the secrecy uploading rate constraint $\eqref{eq:18d}$, respectively.

\subsection{TD3-Based Solution }

Taking into account the high-dimensional state and action space in the RIS-assisted FL environment, we adopt the TD3-based PSBD algorithm to tackle the MDP mentioned above. The proposed TD3-based algorithm effectively addresses the issues of Q-value overestimation and model instability commonly found in traditional DRL algorithms [17].
Specifically, the TD3 framework comprises six neural networks including an actor network $\mu\left(s_t\mid\theta^{\mu}\right)$, a target actor network $\mu^{\prime}\left({s_{t}}\mid{\theta^{\mu^{\prime}}}\right)$, two critic networks $Q_{1}\left(s_{t},a_{t}\mid{\theta^{\omega_{1}}}\right)$ and $Q_{2}\left(s_{t},a_{t}\mid{\theta^{\omega_{2}}}\right)$,
and two target critic networks $Q_{1}^{\prime}\left(s_{t},a_{t}\mid{\theta^{\omega_{1}^{\prime}}}\right)$ and $Q_{2}^{\prime}\left(s_{t},a_{t}\mid{\theta^{\omega_{2}^{\prime}}}\right)$, where $\theta^{\mu}$, $\theta^{\mu^{\prime}}$, $\theta^{\omega_{1}}$, $\theta^{\omega_{2}}$, $\theta^{\omega_{1}^{\prime}}$, and $\theta^{\omega_{2}^{\prime}}$ are weight parameters in the six networks, respectively. The actor network is tasked with learning a mapping function that approximates optimal policies and producing actions. The two critic networks independently evaluate the value of the actions taken by the actor network, allowing for a more robust estimation of Q-values. Additionally, the target networks contribute to further stabilizing the overall training process.

At each time step $t$, the agent gets a current state $s_t$ from the environment and the actor network selects an
action $a_t$ from action space based on $s_t$. To poise the exploration of new actions and the exploitation of known actions, a random noise  $\epsilon$ is appended to the action as $a_{t}=\mu\left(s_{t}\mid\theta^{\mu}\right)+\epsilon$, where  $\epsilon\sim\mathcal{N}\left(0,\sigma^{2}\right)$ is the noise with zero mean and variance  $\sigma^{2}$. Then, once the action is executed, a corresponding reward $r_t$ can be acquired, and $s_t$ will evolve into a next state $s_{t+1}$. Through this interactive process, the agent acquires a tuple of $(s_t,a_t,r_t,s_{t+1})$ and stores it into an experience replay buffer $\mathcal{D}$.  In each time step, minibatch tuples of ${(s_i,a_i,r_i,s_{i+1})}_{i=1}^I$ are randomly sampled from $\mathcal{D}$ to train the networks, where $i$ denotes the index of each tuple and $I$ represents the number of tuples.

By randomly sampling a tuple $(s_i,a_i,r_i,s_{i+1})$ from the experience replay buffer, the two critic networks are updated based on the loss function that measures the difference between  the target Q-value and the estimated Q-value, thereby optimizing the accuracy of the Q-function, the loss function is formulated as

\begin{equation}
    L\!\left({\theta^{\omega_j}}\right)\!{=}\!\frac{1}{I}\!\sum_{i=1}^{I}\!\left[y_i\!-\!Q\!_j\!\left(s_i,\!a_i\!\mid\!{\theta^{\omega_j}}\right)\right]^2\!, \,\, \text{for} \,{j{=}1,2},
\end{equation}
where $j$ means the index of the two critic networks and $y_i$ represents the approximate target Q-value that can be calculated as

 \begin{equation}
    y_i=r\left(s_i,a_i\right)+\gamma\min_{j=1,2}Q_j^{\prime}\left(s_{i},a_{i}\mid{\theta^{\omega_j^{\prime}}}\right),
\end{equation}
where $\gamma$ stands for the reward discount factor. When the critic networks are updated, the target Q-value produced by the deterministic policies is susceptible to the over-fitting of Q-value caused by the Q-function estimation errors \cite{r_18}. TD3 algorithm introduces a concept of target policy smoothing, which is used to avoid the over-fitting of Q-value by adding a clipped Gaussian noise $\epsilon^{\prime}$ to the target actions as

\begin{equation}
    a_{i+1}\left(s_{i+1}\right)={\pi}^{\prime}\left(s_{i+1}\mid{\mu}^{\prime}\right)+\epsilon^{\prime},
\end{equation}
where $\epsilon^{\prime}\sim\mathrm{clip}\left(\mathcal{N}\left(0,\sigma_a^{2}\right),-c,c\right)$ is the noise that follows the standard normal distribution with zero mean and variance $\sigma_a^2$ and $\mathrm{clip}(\cdot)$ means the clipping function that limits the noise within the range of $[-c,c]$.

Then, the parameters $\theta^{\omega_j}$ of the critic networks are updated by gradient descent as

\begin{equation}
\begin{aligned}
    \nabla_{{\theta^{\omega_{j}}}}L\left({\theta^{\omega_{j}}}\right) &{=} {\frac{1}{I}}                                                      \!\sum_{i=1}^{I}\!\left(y_{i}{ -} Q_{j}\left(s_{i},a_{i}{\mid}{\theta^{\omega_{j}}}\right)\right)  \\ &\nabla_{\theta^{\omega_{j}}}Q_{j}\left(s_{i},a_{i}{\mid}{\theta^{\omega_{j}}}\right),
\end{aligned}
\end{equation}

    \begin{equation}
    {\theta^{\omega_{j}}}\leftarrow{\theta^{\omega_{j}}}-\alpha\nabla_{{\theta^{\omega_{j}}}}L\left({\theta^{\omega_{j}}}\right),
    \end{equation}
    where $\alpha$ indicates the learning rate of the critic networks, $\nabla_{{\theta^{\omega_{j}}}}L\left({\theta^{\omega_{j}}}\right)$ refers to the gradient of the loss function with respect to ${\theta^{\omega_{j}}}$, and $\nabla_{{\theta^{\omega_{j}}}}Q_j(s_t,a_t\mid{\theta^{\omega_{j}}})$ represents the gradient of the Q-value.

To enhance the stability of the training process, the actor network is designed to be updated at a lower frequency than that of the critic networks \cite{r_19}. The main objective of the actor network is to maximize cumulative rewards. To achieve this, the actor network is updated by utilizing the policy gradient method for training as

\begin{equation}
\begin{aligned}
\nabla_{\theta^{\mu}} J(\theta^{\mu}) &= \mathbb{E} \left[ \nabla_{\theta^{\mu}} \mu(s_t \mid \theta^{\mu}) \nabla_{\theta^{\omega_j}} Q_j(s_i, a_i \mid \theta^{\omega_j}) \right] \\
&{=} \frac{1}{I}\!\sum_{i=1}^{I}\!\nabla_{\theta^{\mu}} \mu(s_t \!\mid\! \theta^{\mu}) \nabla_{\theta^{\omega_j}} Q_j(s_i, a_i \!\mid\! \theta^{\omega_j}),
\end{aligned}
\end{equation}
\begin{equation}
\theta^{\mu}\leftarrow\theta^{\mu}+\beta\nabla_{\theta^{\mu}}J\left(\theta^{\mu}\right),
\end{equation}
 where $\beta$ denotes the learning rate of the actor network and
$\nabla_{\theta^{\mu}}\mu(s_i\mid\theta^{\mu})$ equals the gradient of the policy $\mu\left(s_{i}\mid\theta^{\mu}\right)$ with respect to ${\theta^{\mu}}$.

Furthermore, with the aim of guaranteeing stability, the parameters $\theta^{\omega_j^{\prime}}$ and $\theta^{\mu^{\prime}}$ of the three target networks are updated by the soft update mechanism as \cite{r_20}

\begin{equation}
    \theta^{\omega_j^{\prime}}\leftarrow\tau\theta^{\omega_j}+(1-\tau)\theta^{\omega_j^{\prime}},
    \end{equation}
\begin{equation}
    \theta^{\mu^{\prime}}\leftarrow\tau{\theta^\mu}+(1-\tau)\theta^{\mu^{\prime}},
    \end{equation}
where $\tau$ signifies the soft updating rate.

\begin{figure*}[t]
\centering
\subfigure
{\includegraphics[width=0.31\textwidth]{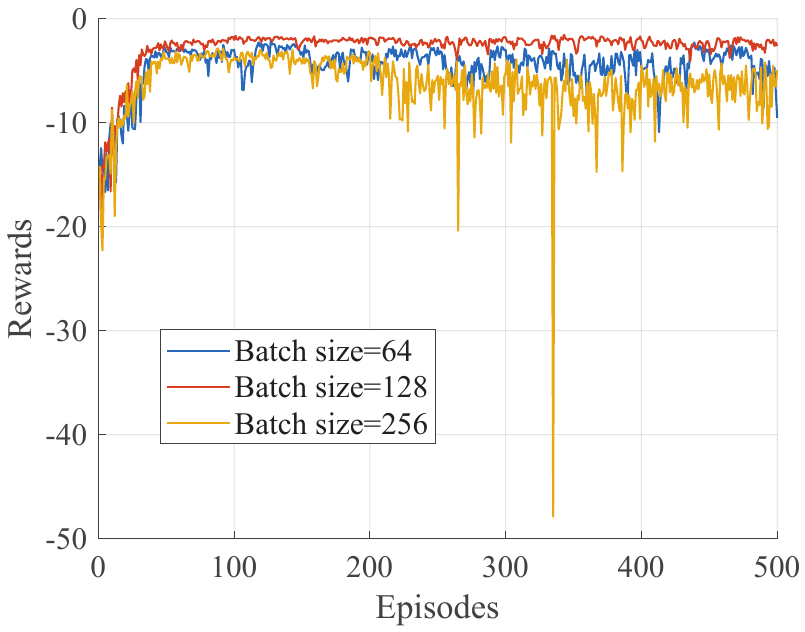}
\label{fig2a}}
\subfigure
{\includegraphics[width=0.32\textwidth]{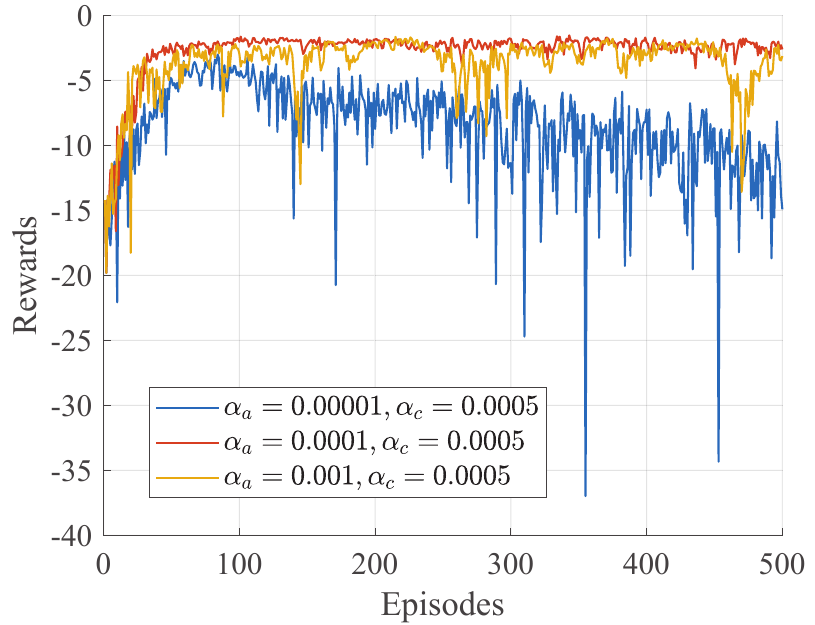}
\label{fig2b}}
\subfigure
{\includegraphics[width=0.31\textwidth]{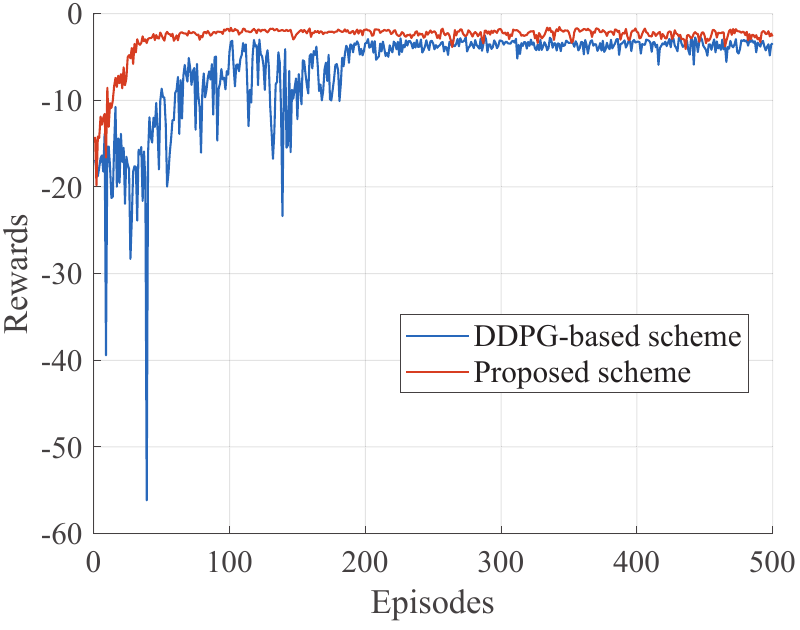}
\label{fig2c}}
\caption{Overall results: (a) Rewards achieved by TD3 w.r.t. training episodes under different batch sizes. (b) Rewards achieved by TD3 w.r.t. training episodes under different learning rates. (c) Rewards achieved by TD3 and DDPG w.r.t. training episodes.
}
\label{fig2}
\end{figure*}

\section{Simulation Results}

\subsection{ Simulation Setup}

We perform simulations to evaluate the performance of the proposed accuracy and security-guaranteed FL scheme. The simulations are conducted in a square area of 60 m $\times$ 60 m. We consider a scenario in which five devices and an eavesdropper are distributed around a BS and a RIS is positioned between the BS and these devices.
Besides, both small-scale fading and large-scale path loss are considered. Thus, wireless channels are modeled as

\begin{equation}
\mathbf{h}_{p,q}\!=\!\sqrt{\varepsilon d_{p,q}^{-\beta_{p,q}}}\!\left(\!\sqrt{\frac{\kappa_{p,q}}{\kappa_{p,q}\!+\!1}}\mathbf{h}_{p,q}^{\mathrm{LoS}}\!+\!\sqrt{\frac{1}{\kappa_{p,q}\!+\!1}}\mathbf{h}_{p,q}^{\mathrm{NLoS}}\!\right),
\end{equation}
where $(p,q){\in}\{{(r,b)},{(\hat{k},r)},{(\bar{k},r)},{(\hat{k},b)},{(\bar{k},b)},{(r,e)},\\{(\hat{k},e)},{(\bar{k},e)}\}$, $\varepsilon$ means the reference path loss at 1m, $d_{p,q}$ is the distance between node ${p}$ and node ${q}$, $\beta_{p,q}$ represents the path loss, $\kappa_{p,q}$ denotes the Rician factor related to the small-scale fading, and $\mathbf{h}_{p,q}^{\mathrm{LoS}}$ and $\mathbf{h}_{p,q}^{\mathrm{NLoS}}$ indicate line-of-sight (LoS) and non-LoS (NLoS) components, respectively.
Following [3], [5] and  [13], we consider the following configurations: $M{=}50$, $S_{\hat k}{=}3$ Mbit, $f_{\hat k}{=}1$ GHz, $c_{\hat k}{=}1000$ cycles/bit, ${D_{\hat k}}{=}6250$ bits, ${P_{\hat k}}{=}{P_{\bar k}}{=}0.1$ W, $\sigma_b^2{=}\sigma_e^2{=}10^{-14}$ W, $R_{\hat k}^{\min}{=}2$, $10^4$ bps, $\varepsilon{=}-30$ dB, $\beta_{k,r}{=}\beta_{r,b}{=}\beta_{r,e}{=}2.2$, $\beta_{k,b}{=}\beta_{k,e}{=}3.6$, $\kappa_{r,k}{=}\kappa_{r,b}{=}\kappa_{r,e}{=}4$, and $\kappa_{k,b}{=}\kappa_{k,e}{=}0$.

We consider the proposed TD3-based algorithm is implemented using PyTorch 1.12.0 in Python 3.8.8. Also, the actor network comprises two hidden layers with neuron sizes of 64 and 64, and each critic network contains two hidden layers with neuron sizes of 512 and 512. The learning rates, the discount factor, the soft update coefficient, the size of the mini-batches, and the size of the experience replay buffer are configured as $\alpha_a$=0.0001, $\alpha_c$=0.0005, ${\gamma}$ = 0.99, ${\tau}$=0.001, ${I}$= 128, and  $\left| \mathcal{D} \right| =$ 10000, respectively.

\subsection{Analysis of Simulation Results}

We present the convergence behaviors of our proposed algorithm in Fig. 2. Specifically, Fig. 2(a) shows the rewards achieved by the proposed TD3 algorithm with respect to (w.r.t.) training episodes under different batch sizes.  Compared to batch sizes 64 and 256, the proposed TD3-based algorithm achieves better performance when the batch size is 128, while convergence performance under the other two batch sizes oscillates greatly. Fig. 2(b) illustrates the rewards obtained by the proposed TD3 algorithm versus training episodes under various learning rates. We can observe that the TD3 algorithm with ${\alpha_a}$=0.0001 and ${\alpha_c}$ = 0.0005 exhibits higher rewards. Fig. 2(c) compares the rewards achieved by both TD3 and the DDPG algorithm concerning training episodes. It can be observed that TD3 yields higher rewards than DDPG. Also, the rewards of TD3 eventually converge to a stable value after approximately 50 training episodes, while DDPG converges to a stable value after 200 training episodes. This result illustrates the superiority of our proposed algorithm in terms of convergence.

To further demonstrate the effectiveness of the proposed scheme, we then compare it with the following four baselines:
\begin{itemize}
\item\textit{Fixed bandwidth allocation (FBA):}Fixed bandwidth resources are allocated to participants for FL, while ${x_{k}}$ and $\mathbf{\Theta}$ are jointly optimized to minimize the training latency of FL.
\item\textit{Random device selection (RDS): }Participants are randomly selected to take part in FL. Meanwhile, jammers can be randomly selected to prevent eavesdropping.
\item\textit{Proposed scheme with random RIS phase:}This scheme randomly generates RIS phase shifts for uploading model parameters.
\item\textit{DDPG-based scheme: }The optimization problem in $\eqref{eq:main}$ is addressed by the DDPG algorithm.
\end{itemize}

\begin{figure*}[t]
\centering
\subfigure
{\includegraphics[width=0.32\textwidth]{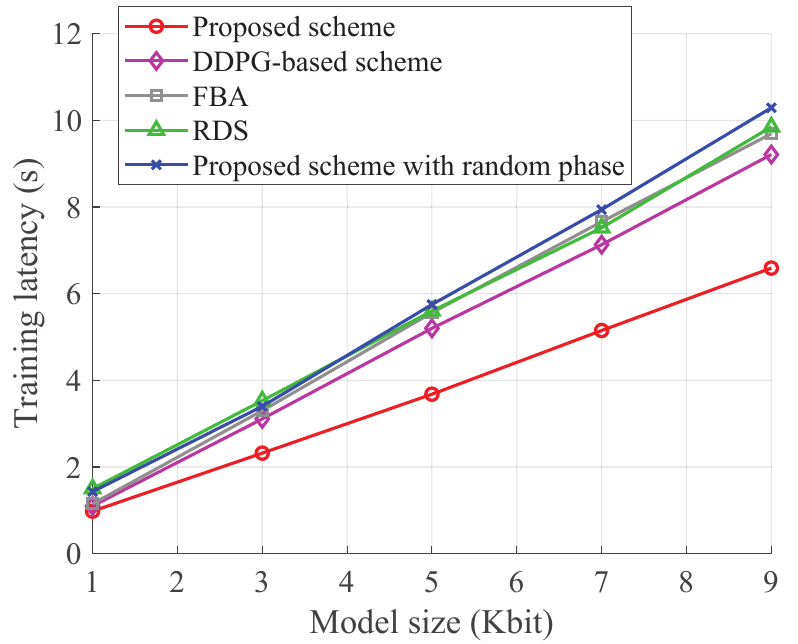}
\label{fig3a}}
\subfigure
{\includegraphics[width=0.32\textwidth]{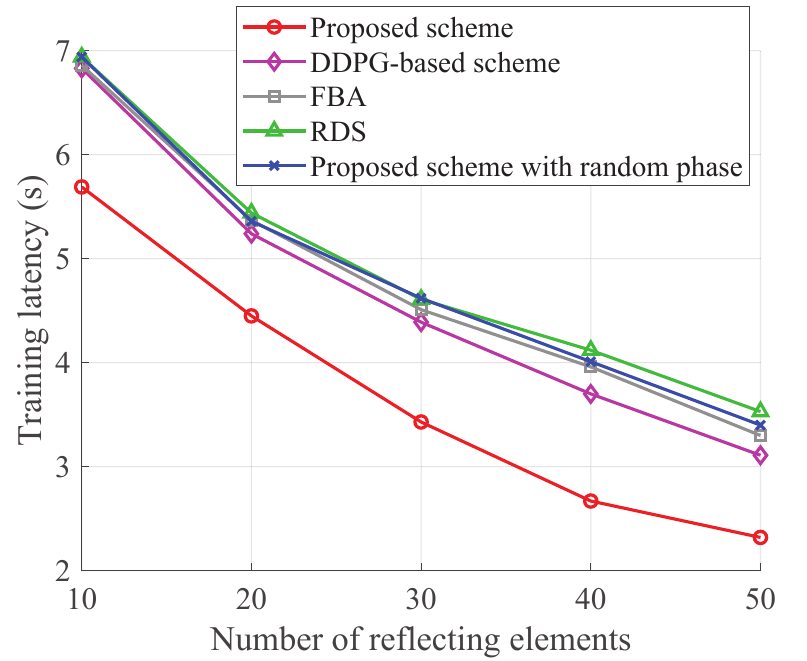}
\label{fig3b}}
\subfigure
{\includegraphics[width=0.32\textwidth]{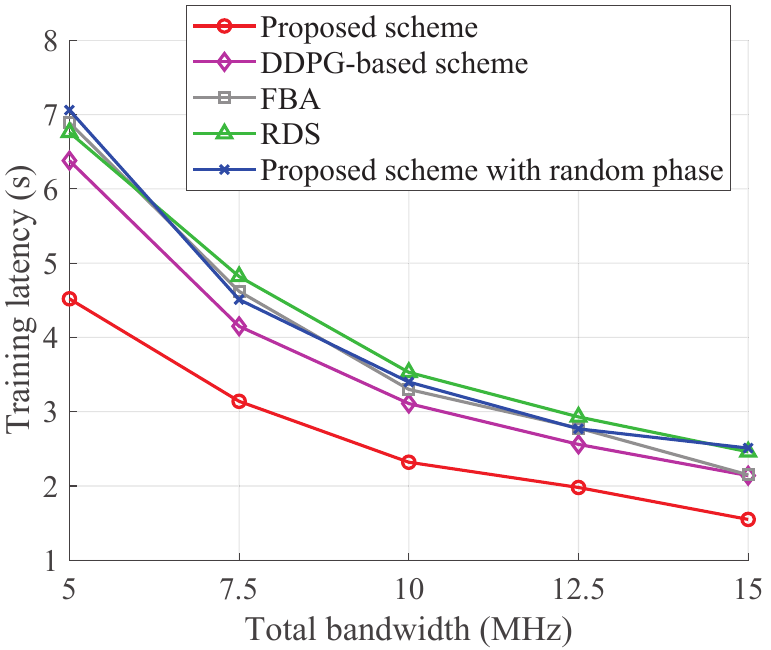}
\label{fig3c}}
\caption{Performance comparison: (a) Training latency w.r.t. model size. (b) Training latency w.r.t. the number of RIS reflecting elements. (c) Training latency w.r.t. total bandwidth.
}
\label{fig3}
\end{figure*}

Fig. 3(a) illustrates the comparison of the training latency of the proposed scheme  and the baselines under different model sizes. It is obvious that as the model size increases, the total latency also rises.
This trend can be attributed to the fact that a larger amount of local model parameters needs to be uploaded, which results in increasing model uploading latency.Moreover, the proposed scheme consistently  outperforms the baselines, which indicates its ability to reduce the training latency. This advantage stems from the joint optimization of participant selection, bandwidth allocation, and beamforming design at the RIS in the proposed scheme.

We show the simulation results of training latency against different numbers of reflecting elements between the proposed scheme and the baselines in Fig. 3(b). We can evidently observe that the training latency demonstrates a decreasing trend when the number of RIS reflecting elements increases. The reason is that increasing the number of
RIS reflecting elements can offer enhanced spatial freedom, resulting in higher channel gains for improving system performance. The results indicate that deploying an adequate number of reflecting elements in the system holds the potential to reduce training latency.

Fig. 3(c) depicts the impact of total bandwidth on the latency of a global training round for FL. As the total bandwidth increases from 5 to 15 MHz, the training latency in all the schemes decreases. This reduction can be ascribed to the increase in the achievable rate from edge devices to the BS, consequently reducing the latency in uploading local model parameters. Again, the proposed scheme shows superior performance compared to the baselines. Particularly, the performance improvements of the proposed scheme are approximately 25$\%$, 31$\%$, 34$\%$, and 33$\%$ compared to DDPG, FBA, RDS, and proposed scheme with random phase, respectively. This emphasizes the effectiveness of TD3 in optimizing the variables in $\eqref{eq:main}$ and underscores its capability to enhance system performance. Also, the RDS scheme exhibits the worst performance, which highlights the significant impact of participant selection on the training latency.

\begin{figure}[htbp]
\centering
\includegraphics[width=0.48\textwidth]{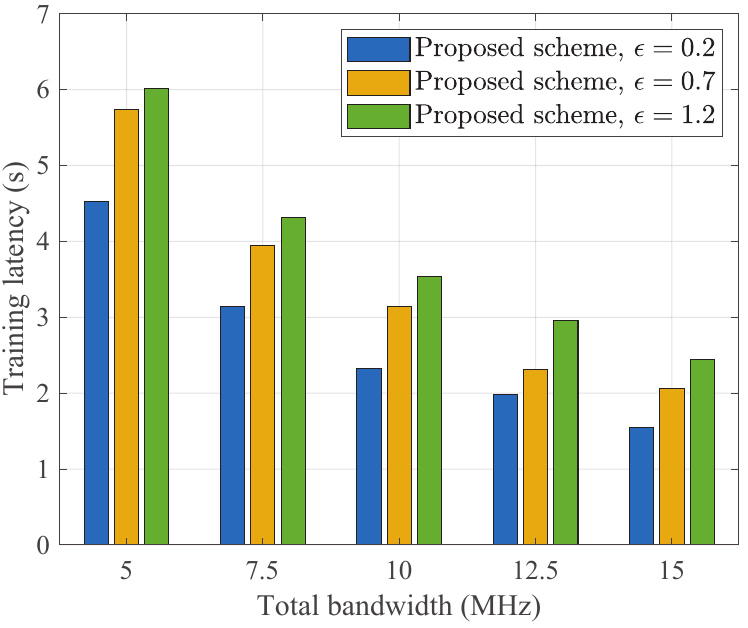}
\caption{Training latency w.r.t. total bandwidth under different $\epsilon$.}
\label{fig4}
\end{figure}

Figure 4 shows the training latency of FL w.r.t. total bandwidth under different convergence accuracy $\epsilon$. The result in Figure~\ref{fig4} demonstrates that the convergence accuracy $\epsilon$ affects the training latency. We can observe that training latency increases as $\epsilon$ increases. It can be attributed to the fact that the number of participants is negatively correlated with $\epsilon$ according to (17c).
When $\epsilon$ is lower, more participants are selected to participate in the training process, which results in reducing channel gains of jamming links. Also, this reduction can increase the achievable rate for uploading local model parameters.

\section{conclution}

In this paper, we have studied accuracy and security-guaranteed participant selection and beamforming design for a RIS-assisted FL system. In this system, a subset of edge devices with powerful computational capabilities or good communication conditions are selected as participants in the training process. Meanwhile, the remaining devices act as cooperative jammers to interfere with eavesdropping. We have formulated a latency minimization problem while meeting the convergence and security requirements. To address this challenging MINLP problem, we have proposed a TD3 algorithm to determine participant selection, bandwidth allocation, and RIS beamforming design. Simulation results have verified the effectiveness of our proposed scheme.

\appendix
\subsection{Proof of Theorem 1}
\label{app:a}

Based on Assumption 1, we can reexpress the second-order Taylor expansion of
$F({\textbf{\textit{w}}^{t + 1}})$ as

\begin{equation}
\tag{A.1}
			\begin{array}{l}
				\!\!\!F({\textbf{\textit{w}}^{t + 1}}){ \le} F({\textbf{\textit{w}}^t}) {+} {(\!{\textbf{\textit{w}}^{t + 1}}{ -} {\textbf{\textit{w}}^t})^T}\!\nabla\! F({\textbf{\textit{w}}^t}) {+ }\frac{\mu }{2}{\left\| {{\textbf{\textit{w}}^{t + 1}}{ - }{\textbf{\textit{w}}^t}} \right\|^2}\\
				\!\!\!{=} F({\textbf{\textit{w}}^t}) {-} \alpha {(\!                                                        \nabla \!F({\textbf{\textit{w}}^t}){ -} \chi )^T}\nabla \!F({\textbf{\textit{w}}^t}){ +} \frac{{\mu {\alpha ^2}}}{2}{\left\| {\nabla\! F({\textbf{\textit{w}}^t}){ -} \chi } \right\|^2}\!,
			\end{array}
		\end{equation}
		where
		$\chi  = \nabla F({\textbf{\textit{w}}^t}) - {g^t}$.
Given $\alpha  = \frac{1}{\mu }$, we can derive

\begin{equation}
\tag{A.2}
			\!\!F({\textbf{\textit{w}}^{t + 1}}) \le F({\textbf{\textit{w}}^t}){-} \frac{1}{{2\mu }}{\left\| {\nabla F({\textbf{\textit{w}}^t})} \right\|^2} + \frac{1}{{2\mu }}{\left\| \chi  \right\|^2}.
		\end{equation}
By performing expectation operations on both sides of (A.2), we can obtain

\begin{equation}
\tag{A.3}
			\!\mathbb{E}[\!F({\textbf{\textit{w}}^{t + 1}})] {\le} \mathbb{E}[\!F({\textbf\!{\textit{w}}^t})] \!{-} \frac{1}{{2\mu }}\mathbb{E}[{\left\| {\nabla\! F({\textbf{\textit{w}}^t})} \right\|\!\!^2}] {+} \frac{1}{{2\mu }}\mathbb{E}[{\left\| \chi  \right\|\!^2}],
		\end{equation}
		where $\mathbb{E}[{\left\| \chi  \right\|^2}] = \mathbb{E}[{\left\| {\nabla F({\textbf{\textit{w}}^t}) - {\textbf{\textit{g}}^{t}}} \right\|^2}]$.
		
		Next, we introduce an auxiliary variable
		$\hat {\textbf{\textit{g}}}{ =} \frac{1}{K}\sum\limits_{k = 1}^K {x_k{{\textbf{\textit{g}}}_k^t} = \frac{{\sum\limits_{k = 1}^K {x_k\frac{1}{{{D_k}}}\sum\limits_{i = 1}^{{D_k}} {\nabla f({\textbf{\textit{w}}_k^t};{\textbf{\textit{u}}_{ki}},{\textit{v}_{ki}})} } }}{K}} $. Based on $\eqref{eq:1}$, $\eqref{eq:2}$, and $\eqref{eq:4}$, we have

		\begin{equation}
\tag{A.4}\label{eq:43}
			\nabla F(\textbf{\textit{w}}^t) = \frac{{\sum\limits_{k = 1}^K {\frac{1}{{{D_k}}}\sum\limits_{i = 1}^{{D_k}} {\nabla f({\textbf{\textit{w}}_k^t};{\textbf{\textit{u}}_{ki}},{\textit{v}_{ki}})}}}}{K},
		\end{equation}

		\begin{equation}
\tag{A.5}\label{eq:44}
			{\textbf{\textit{g}}^t} = \frac{{\sum\limits_{k = 1}^K {\frac{{{x_k}}}{{{D_k}}}\sum\limits_{i = 1}^{{D_k}} {\nabla f({\textbf{\textit{w}}_k^t};{\textbf{\textit{u}}_{ki}},{\textit{v}_{ki}})} } }}{{\left| {{{\cal A}}} \right|}},
		\end{equation}
where $\textbf{\textit{g}}^t$ is the global gradient.	
	
		It is konwn that
		${\left\| {X + Y} \right\|^2} = {\left\| X \right\|^2} + 2\left\langle {X,Y} \right\rangle  + {\left\| Y \right\|^2}$. From Young's inequality
		$\left\langle {X,Y} \right\rangle  \le \frac{1}{2}{\left\| X \right\|^2} + \frac{1}{2}{\left\| Y \right\|^2}$, we can derive
		${\left\| {X + Y} \right\|^2} \le 2({\left\| X \right\|^2} + {\left\| Y \right\|^2})$. Therefore, we can acquire  the following inequality:

		\begin{equation}
\tag{A.6}
			\begin{array}{l}
				\mathbb{E}[{\left\| {\nabla F(\textbf{\textit{w}}^t) - {\textbf{\textit{g}}^t}} \right\|^2}] = \mathbb{E}[{\left\| {\nabla F(\textbf{\textit{w}}^t) - \hat {\textbf{\textit{g}}} + \hat {\textbf{\textit{g}}} - {\textbf{\textit{g}}^t}} \right\|^2}]\\ \le 2\mathbb{E}[{\left\| {\nabla F(\textbf{\textit{w}}^t) - \hat {\textbf{\textit{g}}}} \right\|^2}] + 2\mathbb{E}[{\left\| {\hat {\textbf{\textit{g}}} - {\textbf{\textit{g}}^t}} \right\|^2}]\\
				= 2\mathbb{E}[{\left\| {\frac{{\sum\limits_{k = 1}^K {(1 - x_k){\textbf{\textit{g}}_k^t}} }}{K}} \right\|^2}] + 2\mathbb{E}[{\left\| {\frac{{(\left| {{{\cal A}}} \right| - K)\sum\limits_{k = 1}^K {x_k{\textbf{\textit{g}}_k^t}} }}{{K\left| {{{\cal A}_t}} \right|}}} \right\|^2}]\\
				= \frac{2}{{{K^2}}}\!\!\sum\limits_{k = 1}^K \!{(1 {-} x_k)} \mathbb{E}[{\left\| {{\textbf{\textit{g}}_k^t}} \right\|^2}]{ +} \frac{{2{{(\left| {{{\cal A}}} \right| - K)}^2}}}{{{K^2}{{\left| {{{\cal A}}} \right|}^2}}}\!\sum\limits_{k = 1}^K {x_k\mathbb{E}[{{\left\| {{\textbf{\textit{g}}_k^t}} \right\|}^2}]}.
			\end{array}
		\end{equation}
		According to Assumption 4, it can be inferred that

\begin{equation}
\tag{A.7}
\begin{aligned}
\mathbb{E}[{\left\| \chi  \right\|^2}]
&\le \frac{2\delta}{K}\sum_{k=1}^K (1 - x_k) + \frac{2\delta (\left| \mathcal{A} \right| - K)^2}{K \left| \mathcal{A} \right|^2} \sum_{k=1}^K x_k \\
&= 2\delta + \frac{2\delta (K^2 - 2\left| \mathcal{A} \right|)}{\left| \mathcal{A} \right|^2} \sum_{k=1}^K x_k,
\end{aligned}
\end{equation}

		\begin{equation}
\tag{A.8}
			\begin{aligned}
				\mathbb{E}[{\left\|{\nabla\! F\!({\textbf{\textit{w}}^t})} \right\|^2}]
&\le 2\mu \mathbb{E}[F({\textbf{\textit{w}}^t})] \\
 &-2\mu \mathbb{E}[F({\textbf{\textit{w}}^{t + 1}})] + \mathbb{E}[{\left\| \chi  \right\|^2}].
\end{aligned}
		\end{equation}
		By summing up the inequalities in (A.5) and (A.6) from $t=0$ to $t=\Omega $, we obtain

		\begin{equation}
\tag{A.9}
			\begin{array}{l}
				\frac{1}{{\Omega  + 1}}\sum\limits_{t = 0}^\Omega  \mathbb{E}{[{{\left\| {\nabla F({\textbf{\textit{w}}^t})} \right\|}^2}]}  \le \frac{{2\mu \sum\limits_{t = 0}^\Omega  {\{\mathbb{E} [F({\textbf{\textit{w}}^t})] - \mathbb{E}[F({\textbf{\textit{w}}^{t + 1}})]\} } }}{{\Omega  + 1}} + \\
				\frac{{2\delta }}{{K(\Omega  + 1)}}\sum\limits_{t = 0}^\Omega  {\sum\limits_{k = 1}^K {(1 - x_k)} }  + \frac{{2\delta {{(\left| {{{\cal A}}} \right| - K)}^2}}}{{K{{\left| {{{\cal A}}} \right|}^2}(\Omega  + 1)}}\sum\limits_{t = 0}^\Omega  {\sum\limits_{k = 1}^K {x_k} } \\
				\le \frac{{2\mu (F({\textbf{\textit{w}}^0}) - {F^*})}}{{\Omega  + 1}} + 2\delta  + \frac{{2\delta ({K} - 2\left| {{{\cal A}}} \right|)}}{{{{\left| {{{\cal A}}} \right|}^2}}} {\sum\limits_{k = 1}^K {x_k} }.
			\end{array}
		\end{equation}
		This concludes the proof of Theorem 1.

\ifCLASSOPTIONcaptionsoff
  \newpage
\fi

\end{document}